\documentclass[pra,letterpaper,onecolumn,showpacs,superscriptaddress,floatfix]{revtex4}

\usepackage{graphicx,psfrag,amsmath,amssymb,amsfonts,bbm,latexsym,color,dcolumn,bm}

\begin{document}

\title{Effective microscopic models for sympathetic cooling of atomic gases}

\author{Roberto Onofrio}

\affiliation{\mbox{Dipartimento di Fisica e Astronomia ``Galileo Galilei'', Universit\`a  di Padova, 
Via Marzolo 8, Padova 35131, Italy}}

\affiliation{\mbox{Department of Physics and Astronomy, Dartmouth College, 6127 Wilder Laboratory, 
Hanover, NH 03755, USA}}

\author{Bala Sundaram}

\affiliation{\mbox{Department of Physics, University of Massachusetts, Boston, MA 02125, USA}}

\begin{abstract}
Thermalization of a system in the presence of a heat bath has been the subject of 
many theoretical investigations especially in the framework of solid state physics. 
In this setting, the presence of a large bandwidth for the frequency distribution of the harmonic 
oscillators schematizing the heat bath is crucial, as emphasized in the Caldeira-Leggett model. 
By contrast, ultracold gases in atomic traps oscillate at well-defined frequencies and therefore 
seem to lie outside the Caldeira-Leggett paradigm. We introduce interaction Hamiltonians which 
allow us to adapt the model to an atomic physics framework.
The intrinsic nonlinearity of these models differentiates them from the original Caldeira-Leggett 
model and calls for a nontrivial stability analysis to determine effective ranges for the model 
parameters. 
These models allow for molecular dynamics simulations of mixtures of ultracold gases, which is
of current relevance for optimizing sympathetic cooling in degenerate Bose-Fermi mixtures.
\end{abstract}


\maketitle

\section{Introduction}
The study of dynamical systems interacting with an external environment plays an essential 
role in classical and quantum physics, since long-term properties of any system are going to 
be affected by the outside world unless complete adiabaticity can be assumed at all times.
In particular, the presence of environments having the energy of their constituent particles 
shared according to the Boltzmann distribution, also named `baths', is at the basis of the canonical 
ensemble approach in equilibrium statistical mechanics, with its large variety of physical implications. 
A microscopic analysis of the system-bath interplay is possible by writing explicitly the 
bath degrees of freedom and integrating out their effect on the target particle. 
Along these lines, various authors have determined sufficient conditions under which a Langevin dynamics, 
and the consequent thermalization, holds. This has been achieved by considering a bath made of 
non-interacting harmonic oscillators,  which interact linearly with the target particle, the 
so-called Caldeira-Leggett model \cite{Magalinskii,Ullersma,Caldeira1,Caldeira2}. 

More recently, studies have focused attention on baths endowed with finite resources, 
such as a finite number of harmonic oscillators distributed in a finite bandwidth, 
{\it i.e.} allowing for a non-zero infrared cut-off and a finite ultraviolet cut-off 
\cite{Taylor,Hanggi,Hasegawa1,Carcaterra,Hasegawa2,Hasegawa3}. 
In particular, the dynamics of a target harmonic oscillator interacting via a linear 
and translationally invariant term with a bath, composed of a finite number of harmonic 
oscillators whose frequencies are all in a limited range, was studied in detail.
The principal outcome of this analysis, presented in \cite{Taylor}, can be summarized as follows. 
To ensure thermalization, the frequency of the test particle must lie within the band 
of bath frequencies, which in turn must also have a large enough bandwidth. 
In the absence of any one of these conditions the test particle does not thermalize to 
the temperature of the bath. The implication of this is either that the energy distribution 
is not amenable to a Boltzmann distribution or that the resulting effective temperature 
of the test particle is significantly different from that of the bath. 

Thermalization plays a crucial role in sympathetic cooling of atomic species trapped in magnetic or 
optical potentials. This was first demonstrated experimentally when  ${}^{85}$Rb, a bosonic Rubidium
isotope which is hard to cool using evaporating techniques, was successfully cooled via thermal contact 
with ${}^{87}$Rb \cite{Ensher}. The technique is even more crucial in the case of fermionic atomic species 
where direct evaporative cooling loses its efficiency in the degenerate regime due to fundamental 
limitations arising from the Pauli principle \cite{DeMarcoJin,Crescimanno,Holland}. 
A number of studies have led to predictions for new phase transitions occurring for ultracold fermions in 
single traps or optical lattices, and the current experimental focus is on achieving temperatures low 
enough to observe these effects \cite{DeMarco}.  

The Caldeira-Leggett model, in its original form, cannot be used to describe thermalization in the 
atomic physics framework. In its applications to solid-state systems, this model is meaningful due 
to the multiplicity of normal modes associated with lattice vibrations and the need for a continuous 
density of states, as in the Debye model \cite{Lepri}. In contrast,  in atom traps no phonon-mediated 
interactions exist in principle either in single traps or in optical lattices. 
Thus coolant and target species both have single values for the trapping frequencies 
which means that, based on the Caldeira-Leggett model, no thermalization is expected. 
This is further heightened in those proposals where the two species are deliberately required 
to have significantly different trapping frequencies in order to optimize sympathetic cooling 
\cite{Onofrio1,Onofrio2,Brown,Catani,Tassy,Baumer,Vaidya}. Of course, this inference is in stark 
contrast to the experimentally observed success of sympathetic cooling in a range of atomic mixtures. 

In this paper, we demonstrate that it is possible to choose bath-test particle interactions which reconcile
the Caldeira-Leggett model with the experimentally manifest effectiveness of sympathetic cooling. 
The details of the dynamics of thermalization is studied as well in terms of the stability with respect 
to different choices of the model parameters. The model we discuss is purely classical, both in 
regard to the dynamical evolution and to the choice of Boltzmann heat baths. While classical dynamics is 
generally consider sufficient to describe the motion of ultracold atoms under current experimental 
conditions, in the quantum degenerate regime the Boltzmann energy distribution should be superseded by 
Bose or Fermi distributions. This work should be therefore considered as a prelude, in the Dulong-Petit 
classical limit, to the analysis of fully degenerate ultracold quantum gases, setting the stage and 
the language for future work aimed at a more comprehensive analysis of dynamical situations such as 
the sympathetic cooling of Bose-Fermi mixtures.

The paper is organized as follows. In Section II we introduce two interaction Hamiltonians 
differing in their selectivity with respect to the relative velocity of the interacting particles, 
and we make explicit the Hamilton's equations and associated considerations. 
In Section III we discuss an interesting phenomenon which occurs in the strong coupling limit of repulsive 
interaction, {\it i.e.} the fact that the test particle experiences an effective double-well potential which 
influences the dynamics of cooling. This is demonstrated both through numerical and approximate fixed point 
analyses, which then results in a phase diagram delineating various regimes in the parameter space. 
Section IV presents the results of numerical simulations showing the approach to thermal equilibrium and 
the relaxation dynamics in terms of the interaction energy. In the process, we revisit some assumptions 
implicit in discussions of thermalization, which are not necessarily satisfied in models like ours. 
In the conclusions, we discuss the relevance of this model to actual simulations of ultracold atomic 
mixtures, as well as several directions for expanding the applicability of our models.
A particular point of interest is the possible application of recent results on shortcuts to 
adiabaticity for hastening the rate of thermalization in our dynamical models.  

\section{Interaction Hamiltonians}

The classical Hamiltonian for a 1D harmonic oscillator of mass $M$ and angular frequency $\Omega$, 
which we will refer to as a 'test particle', in the presence of a bath made of $N_b$ harmonic oscillators 
of mass $m$ and generic frequency $\omega_n \in [ \omega_{IR}, \omega_{UV} ]$ linearly coupled to it, is 
written as

\begin{equation}
H_{tot} = \frac{P^2}{2M} +  \frac{1}{2} M \Omega^2 Q^2 + \sum_{n=1}^{N_b} \left  [  \frac{p_{n}^2}{2m} + 
\frac{1}{2}m \omega_{n}^2 (q_n - Q)^2    \right ].
\end{equation}

The associated Hamilton equations are 
\begin{eqnarray}
\dot{Q} &=& \frac{P}{M} \;, \nonumber  \\
\dot{P} &=& - M\Omega^2 Q + \sum_{n=1}^{N_b} m\omega_n^2(q_n-Q) \;, \nonumber \\
\dot{q_n} &=& \frac{p_n}{m} \;, \nonumber \\
\dot{p_n} &=& -m\omega_n^2 (q_n-Q) \;.
\end{eqnarray}
These show that the effect of the bath on the test particle leads to a modified 'spring constant' 
(which can also be viewed as a combined change in mass and frequency) such that 
$M \Omega^2 \mapsto M \Omega^2+ m \sum_{n=1}^{N_b} \omega_n^2$, and in an effective force equal 
to $-m\sum_{n=1}^{N_b} \omega_n^2 q_n$, where the latter depends on the bath dynamics.
If the bath oscillators all have the same frequency then the driving force is simply proportional 
to $\sum_n q_n$, which averages to zero for a large number of bath oscillators with uniform 
distribution of initial conditions in phase space. This precludes any exchange of net energy 
between the bath and the test particle. This implies that thermalization relies on and requires 
the existence of a large number of distinct frequencies for the bath oscillators. 
As such, atoms oscillating in a harmonic trap with a common angular frequency, $\omega$, do not 
fall within the Caldeira-Leggett framework. 
To address thermalization in this specific situation, the interaction between the test particle and 
the bath particles needs to be modified, and we will consider a Hamiltonian of the generic form

\begin{equation}
H_{tot} = \frac{P^2}{2M} +  \frac{1}{2} M \Omega^2 Q^2 + \sum_{n=1}^{N_b} \left(  \frac{p_{n}^2}{2m} + 
\frac{1}{2}m \omega^2 q_n^2    \right)+ H_{\mathrm{int}}(Q,P,q_n,p_n).
\end{equation}

Given our motivation, the first distinctive requirement in $H_{\mathrm{int}}$ is that interactions between 
atoms be highly localized in configuration space, being described either by pseudopotentials with 
zero range or by finite range (dipolar) interactions. Locality of the bath-test particle interaction, 
first introduced to our knowledge in the context of a microscopic description of a measurement 
apparatus \cite{Namiot}, is easily achieved by introducing a spatial 'filter', for instance of 
a Gaussian nature. This implies an interaction Hamiltonian of the generic form 

\begin{equation}
H_{int}(Q, q_n, P, p_n) = \gamma  \sum_{n=1}^{N_b} f(q_n-Q, v_n-V) \exp \left[-\frac{(q_n-Q)^2}{\lambda^2}\right].
\end{equation}

Thus the test particle-bath interaction Hamiltonian is dependent on two parameters, 
the coupling strength $\gamma$ and the range $\lambda$. The interaction Hamiltonian is 
basically  negligible if $|q_n-Q| >> \lambda$. The generic function $f$ is chosen to fulfill
Galilean invariance, as reflected in the explicit dependence on the differences between 
the coordinates and velocities of the involved particles, with $V=P/M$ and $v_n=p_n/m$. 
For the remainder of our analysis, we will contrast two forms of the function $f$. 
The first is where $f$ is constant, which is a velocity-independent situation, while the 
second choice takes $f$ to be a quadratic function of the relative velocity. 
Having in mind applications in the atomic physics arena, the coupling strength $\gamma$ 
should eventually be related to the scattering length in the case of pseudopotentials.  
Examples of these are the zero-range approximation usually adopted in the Gross-Pitaevskii 
equation for Bose gases (in which case $\lambda$ is chosen to be zero), or to the van der 
Waals potential in the case of finite-range interactions as in dipolar gases.

\subsection{Velocity-independent interaction Hamiltonian}

Of the various possibilities available for the function $f$, the simplest is a constant value, that is
\begin{equation}
H_{\mathrm{int}}(Q, q_n) = \gamma_E  \sum_{n=1}^{N_b} \exp \left[-\frac{(q_n-Q)^2}{\lambda^2}\right] \;.
\end{equation}

\noindent 
Here a significant impulsive force occurs at each interaction between the test particle 
and a particle of the bath if they are within a distance of order of $\lambda$ of each other. 
The strength of the interaction is controlled by the parameter $\gamma_E$ which has the
dimensions of energy and sets the maximal possible energy exchange during the interaction, which 
is achieved in the limit $\lambda \rightarrow 0$, a sort of force 'kick' of constant amplitude.
The sign of $\gamma_E$ determines the attractive ($\gamma_E<0$ ) or repulsive ($\gamma_E>0$) 
character of the interaction, as in the former case the interaction energy is minimized 
at the smallest distances, and this trend is reversed in the latter situation.
 
The corresponding Hamilton equations of motion are 
\begin{eqnarray}
\dot{Q}&=&\frac{P}{M} \;,\\
\dot{P}&=&-M\Omega^2 Q - 2\frac{\gamma_E}{\lambda^2} 
\sum_{n=1}^{N_b}\exp\left[-\frac{(q_n-Q)^2}{\lambda^2}\right](q_n-Q)\;,\\
\dot{q}_n&=&\frac{p_n}{m} \;, \\
\dot{p}_n&=&-m\omega^2 q_n+ 2\frac{\gamma_E}{\lambda^2}
\exp\left[-\frac{(q_n-Q)^2}{\lambda^2}\right](q_n-Q).
\label{velocitindepequations}
\end{eqnarray}

In the limit of large $\lambda$ the total Hamiltonian resembles Eq. (1) with the uniform bath 
frequency Caldeira-Leggett interaction, but with important differences at large $\gamma$ since, 
by expanding the exponential function to second order in $q_n-Q$, 
\begin{equation}
\exp{[-(q_n-Q)^2/\lambda^2]} \approx 1 - (q_n-Q)^2/\lambda^2 \;,
\label{exponentapprox}
\end{equation}
we get, aside from an irrelevant constant $\gamma_E N_b$, the following Hamiltonian

\begin{equation}
H_{tot} \simeq \frac{P^2}{2M} +  \frac{1}{2} M \Omega^2 
\left(1-\frac{2 \gamma_E N_b}{M \Omega^2 \lambda^2}\right) Q^2+ \sum_{n=1}^{N_b} \left[  \frac{p_{n}^2}{2m} + 
\frac{1}{2}m \omega^2 \left(1-\frac{2\gamma_E}{m \omega^2 \lambda^2}\right)q_n^2 \right]+ 
\frac{2\gamma}{\lambda^2} Q \sum_{n=1}^{N_b} q_n \;.
\end{equation}
This shows that both test- and bath- particle frequencies are renormalized  as
$\Omega \rightarrow \Omega [1-2\gamma_E N_b/(M\Omega^2 \lambda^2)]^{1/2}$, and 
$\omega \rightarrow \omega [1-2\gamma_E/(m\omega^2 \lambda^2)]^{1/2}$. 
Additionally, there is the appearance of an external driving force, proportional to the sum of all 
coordinates of the proximal bath particles, which vanishes in the large, uniformly distributed, $N_b$ limit. 
Notice that, even if the bath particles and the test particle are initially chosen with degenerate 
frequencies and same masses, their interaction produces a relative frequency shift. 
Also, for comparable parameters $M \simeq m, \Omega \simeq \omega$, the harmonic trap becomes 
unstable when $\gamma_E > \bar{\gamma}_E = M\Omega^2 \lambda^2/(2 N_b)$, beyond which the local 
effective potential seen by the test particle near the origin is that of an inverted harmonic oscillator. 
The overall effective potential then morphs into an (attractive) quartic potential at larger values of $Q$, 
as shown by further expanding the right hand side of Eq.~(\ref{exponentapprox}) to fourth order in $q_n-Q$. 
Therefore the effective dynamics for the repulsive $\gamma_E>0$ case is one of a double-well for 
$ 0<\bar{\gamma}_E<\gamma_E$. No such instability occurs around the origin for negative values of 
$\gamma_E$, corresponding to attractive interactions.   

In the opposite limit of small $\lambda$, at any given time the bath decouples into a subsystem of
$N_\lambda< N_b$ interacting oscillators, which are within range $\lambda$ of the test particle, 
while the remaining $(N_b-N_\lambda)$ bath particles are irrelevant to the dynamics. 
The average number of oscillators $N_{\lambda}$ of the heat bath interacting at any given time with 
the test particle may be estimated in a stationary regime once we evaluate the oscillation amplitude 
of the generic oscillator $\langle q_n^2 \rangle^{1/2}=(2 \langle E_n \rangle/m\omega^2)^{1/2}$ where  
$\langle E_n \rangle$ is the total energy of the $n^{th}$ particle. 
The equipartition theorem provides the estimate $E_n=K_B T_b$, so that average oscillation amplitude 
of a generic particle of the heat bath is $\langle q_n^2 \rangle^{1/2} = (2 K_B T_b/(m \omega^2))^{1/2}$, and 
its motion will span an interval of coordinate values $[-\langle q_n^2 \rangle^{1/2},\langle 
q_n^2 \rangle^{1/2}]$, 
while the test particle has a interaction range of $2 \lambda$. 
Considering uniform distributions over the ranges, this suggests that the number of oscillators  
$N_\lambda \simeq 2\lambda N_b/\langle q_n^2 \rangle^{1/2}$,  meaning an inverse dependence on the 
temperature of the heat bath. 

\subsection{Velocity-dependent interaction Hamiltonian}

The form assumed above for the Hamiltonian is the simplest one possible and we can, as more appropriate 
for collisional interactions,  incorporate a dependence on the velocities of the test particle and each 
particle in the heat bath. We continue to assume, as in the previous case, a local spatial interaction 
and the interaction Hamiltonian is now given by 
\begin{equation}
H_{int}(Q,P,q_n,p_n) = \gamma_M  \sum_{n=1}^{N_b} \left(\frac{p_n}{m}-\frac{P}{M}\right)^2 
\exp \left[ -\frac{(q_n-Q)^2}{\lambda^2} \right].
\end{equation}

This velocity-selective Hamiltonian now corresponds to a force which depends on the velocity mismatch.
Our choice is reminiscent of the fact that the elastic scattering rate between atoms is proportional 
to the relative velocity between the colliding particles, provided they are in close proximity to feel 
deviations from the original trajectory through van der Waals like forces. In this case the interaction 
strength is dependent upon a parameter $\gamma_M$ which has dimensions of mass.

The corresponding Hamilton equations are somewhat more complex than earlier due to the velocity-dependent 
term
\begin{eqnarray}
\dot{Q}&=&\frac{P}{M}- 2\frac{\gamma_M}{M} \sum_{n=1}^{N_b} 
\left(\frac{p_n}{m}-\frac{P}{M}\right)
\exp\left[-\frac{(q_n-Q)^2}{\lambda^2}\right]\; , \\
\dot{P}&=&-M\Omega^2 Q- 2\frac{\gamma_M}{\lambda^2} \sum_{n=1}^{N_b}
\left(\frac{p_n}{m}-\frac{P}{M}\right)^2
\exp\left[-\frac{(q_n-Q)^2}{\lambda^2}\right](q_n-Q) \; ,\\
\dot{q_n}&=&\frac{p_n}{m}+ 2\frac{\gamma_M}{M}
\left(\frac{p_n}{m}-\frac{P}{M}\right)
\exp\left[-\frac{(q_n-Q)^2}{\lambda^2}\right] \; ,\\
\dot{p_n}&=&-m\omega^2 q_n+2\frac{\gamma_M}{\lambda^2} 
\left(\frac{p_n}{m}-\frac{P}{M}\right)^2
\exp\left[-\frac{(q_n-Q)^2}{\lambda^2}\right](q_n-Q) \;. 
\label{velocdepequations}
\end{eqnarray}
The main distinctive feature of this set of equations is the presence of time-dependent masses 
for both the target particle and the bath particles, which in a coarse graining of the dynamics 
correspond to time-averaged, effective renormalized masses. 
This feature makes a stability analysis of the solutions more delicate, requiring a fixed point 
analysis as detailed in the next section. In terms of the sign of the interaction, $\gamma_M>0$ 
corresponds to repulsive interactions in which minimization of the interaction energy occurs, as 
in the former case, at large particle separation and minimum relative velocity between the two particles. 
Here, instabilities occur in the attractive case of $\gamma_M<0$, as the interaction energy is minimized 
for small distances between the particles and large velocity separation. 

Although the total energy of the overall system is conserved, the presence, on the righthand side of 
Eqs. (13,14), of forces proportional to the square of the velocity difference resembles dissipation-like 
behavior. This should be expected to make the energy exchange between the test particles and the bath 
particles occur on shorter timescales when contrasted with the velocity-independent case where the 
dissipative feature is lacking. We note here that a more generic interaction Hamiltonian of the form
\begin{equation}
H_{int}(Q,P,q_n,p_n) = \gamma  \sum_{n=1}^{N_b}  
\exp \left[ \frac{\left(p_n/m-P/M\right)^2}{\nu^2}-\frac{(q_n-Q)^2}{\lambda^2} \right].
\end{equation}
can be written where $\nu$ is a parameter related to the velocity spread in the particle ensembles.
A Taylor expansion of the velocity-dependent exponential term generates the velocity-independent 
and velocity-dependent interactions respectively at the zeroth and second order in the parameter 
$(p_n/m-P/M)/\nu$, with the identifications of $\gamma_E=\gamma$ and $\gamma_M=\gamma/\nu^2$.
 
In discussing both types of interactions, what is shown is for a single target particle interacting 
with a set of bath oscillators. These relations can be readily extended to the more general situation 
of $N_{p}$ test particle oscillators interacting with $N_b$ oscillators of the heat bath, as in the 
subsequent simulations described in Section IV.
The advantage of this more general situation is that the effective temperature of the test particles can 
be obtained at any time by looking at the energy distribution of the test particle ensemble, rather than 
taking a finite sequence of energies in a given time interval, a procedure followed in \cite{Taylor}. 
This allows for a more accurate tracking of the time evolution of the temperature, and therefore 
of the expected thermalization dynamics. Besides, the interaction of two clouds of particles is more 
appropriate to sympathetic cooling which is an important motivation for this work.

\begin{figure*}[t] 
\begin{center}
\includegraphics[width=0.48\textwidth]{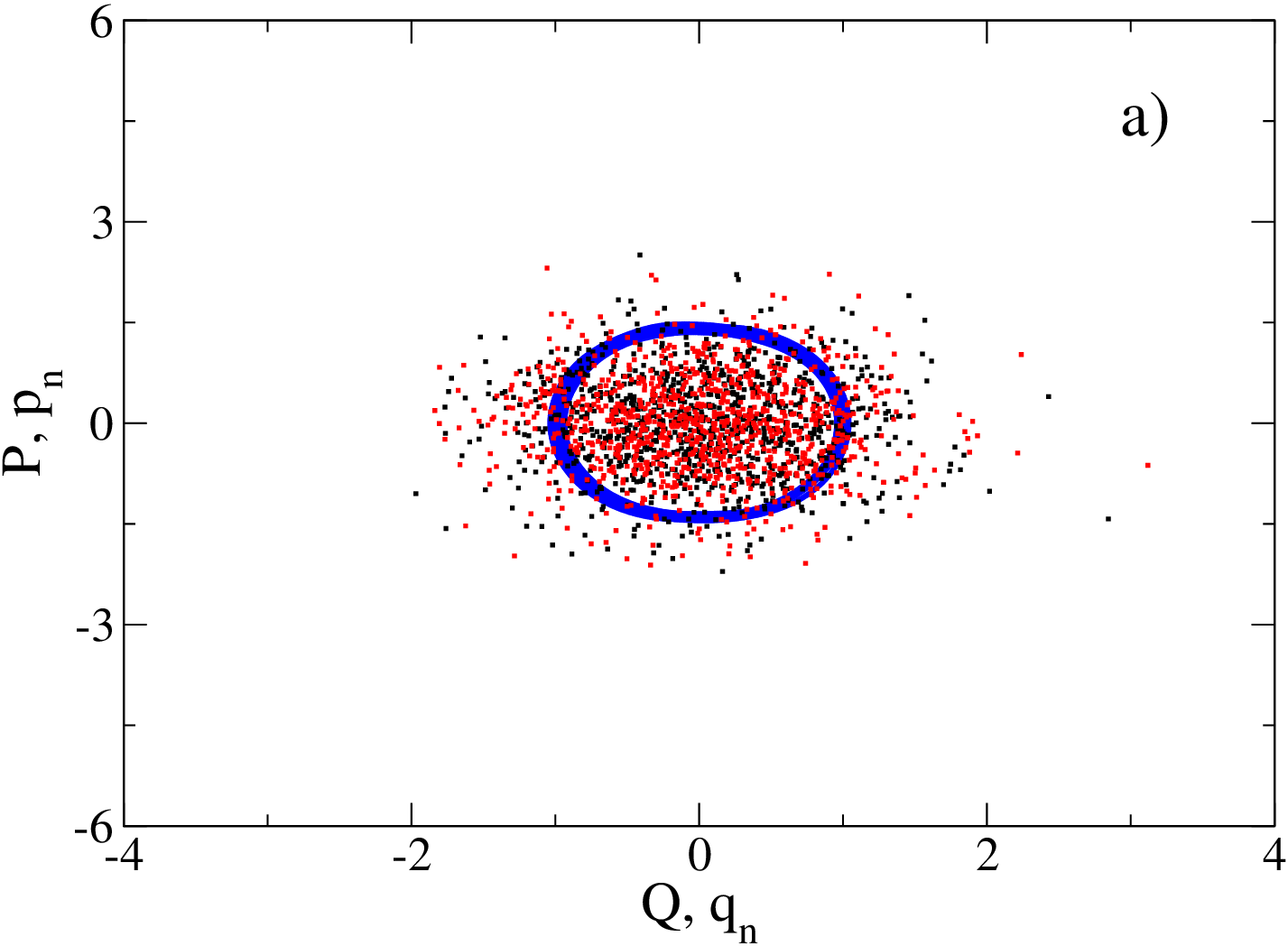}
\hspace{0.2in}
\includegraphics[width=0.48\textwidth]{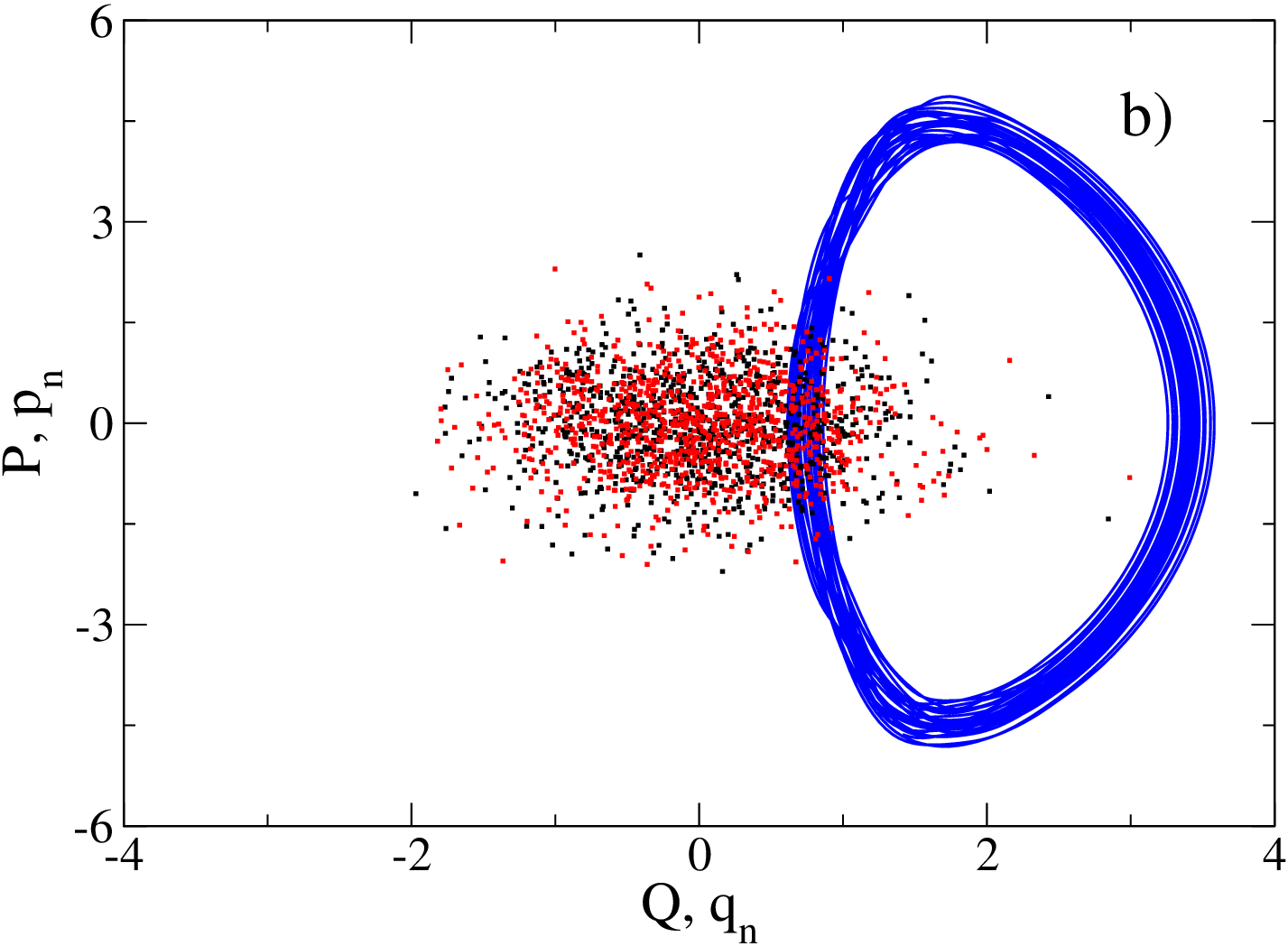}
\end{center}
\caption{(Color online) Demonstration of the transition in the dynamics from (a) a single fixed point 
at $(0,0)$ to (b) ones off-center on changing the parameters of the velocity independent interaction. 
In each of the graphs, the initial (black points) and final (red) phase space locations of the $N_b$ bath 
oscillators are shown as well the blue trajectory of a single test particle. 
Note that the initial condition, chosen here at random, determines the shape of the trajectory in the 
more complicated case (b). Both cases are evaluated for $\lambda=0.25$, one test particle, 
$N_b=10^3$, $T_b=0.5$, $T_{p}=0.1$, $M=m=1$, $\omega_n=1$, and $\Omega=144/89$. 
(a) $\gamma_E=2 \times 10^{-3}$; (b) $\gamma_E=0.1$, with temperatures, masses, angular 
frequencies, parameters $\gamma_E$, $\lambda$ expressed in arbitrary units.}
\label{Fig1}
\end{figure*}

\section{Limits in the parameter space and fixed point analysis}

The analysis thus far, for both forms of the interaction, indicate that there are 
parameter regimes where the thermalization we are interested in may in fact not occur. 
This would limit the range of allowed values of $\gamma$ and $\lambda$.  
More specifically, there is a transition in the test particle motion from the expected one of 
a weakly perturbed harmonic oscillator to an effective double well dynamics as the parameters 
are varied, as shown in Fig. 1 for the case of the velocity-independent interaction model. 
The figures show both the initial and final cloud of bath oscillator locations in phase 
space, which take on the form of Gaussian clouds, and the trajectory of the single target 
particle over a short ($\approx 20$ cycles) time interval. Panel (a) shows the target particle 
trajectory clearly circulating around the origin whereas in case (b) the dynamics is clearly off-center. 
Changing the initial conditions leads to the center of motion switching to a second 
location symmetrically located with respect to the origin of the configurational space. 
To fully understand this phenomenon whose description has been already sketched in the former 
section, we start from the equations of motion for the test particle and look at the conditions 
for the equilibria in the two extreme cases of $\lambda \rightarrow  0$ and of $\lambda$ much 
larger than the confinement size in the trap.

\subsection{Velocity-independent interaction}

We look for fixed points of the Hamilton's equations of motion, by setting $\dot{Q}=0, \dot{P}=0$. 
The equilibrium location $(Q^*, P^*)$ satisfies the conditions
\begin{eqnarray}
P^* &=& 0 \;, \\
M\Omega^2 Q^* & = &- \frac{2\gamma_E}{\lambda^2} \sum_{n=1}^{N_b}  
\exp{\left[-(q_n-Q^*)^2/\lambda^2\right]} (q_n-Q^*) \;.
\end{eqnarray}

We first deal with the case of small $\lambda$, which mimics point-like interactions such as 
those associated with the pseudopotential used in the Gross-Pitaevskii equation, effective mean-field, 
description of a Bose condensate. In this case, $\exp{(-(q_n-Q^*)^2/\lambda^2)} \approx \delta(q_n-Q^*)$ 
which, combined with the $q_n-Q^*$ factor in the righthand side of Eq. (17), implies $Q^*=0$. 
Thus in this limit there is only one fixed point at the origin, $(Q^*, P^*)=(0,0)$. 

The opposite limit of large $\lambda$ can be treated by replacing the exponential 
term with the leading terms of its expansion, which leads to
\begin{equation}
M\Omega^2 Q^* = -\frac{2\gamma_E}{\lambda^2} \sum_n  \left[1-\frac{(q_n-Q^*)^2}{\lambda^2}\right] (q_n-Q^*)= 
\frac{2\gamma_E}{\lambda^2} \left( N_b-3 N_q\right)Q^*-\frac{2\gamma_E N_b}{\lambda^4}{Q^*}^3  \;,
\end{equation}
where we have used the fact that $\sum_n q_n^\alpha$ is assumed to be nonzero only when $\alpha$ is 
even, i.e. a symmetric probability density function $q_n$, and we have introduced the pure number 
$N_q=\sum_{n=1}^{N_b} (q_n/\lambda)^2 << N_b$ in the large $\lambda$ limit we are considering. 
The functional form of this relation is $A(Q^*)^3 = B Q^*$ which means that the possible 
solutions are $Q^*=0$ and $Q^* = \pm \sqrt{B/A}$ with $A= 2 \gamma_E N_b/\lambda^4$
and $B=-M\Omega^2 +2\gamma_E (N_b-3N_q)/\lambda^2$. The threshold for the onset of the 
$Q^* \neq 0$ fixed points is consistent with the analysis following Eq. (10) in the limit 
of $N_b >> N_q$, as they both yield a threshold value of $\bar{\gamma}_E= M \Omega^2 \lambda^2/(2N_b)$.

Moreover, a stability analysis shows that above the threshold  $\bar{\gamma}_E$ the fixed point 
$Q^*=0$ becomes unstable. Given that the fixed point at $(0,0)$ is present both in the small 
as well as large $\lambda$ limits, a fixed point stability analysis around this phase space location 
should be sufficient to gauge the  transition from effective single to double-well behavior 
in the test particle dynamics. This requires the introduction of
\begin{eqnarray}
Q &=& Q^* + \eta = \eta \;, \nonumber \\
P &=& P^* + \epsilon =\epsilon \;,
\end{eqnarray}
leading to the following equations for the small perturbation from the fixed point
\begin{eqnarray}
\dot{\eta} &=& \frac{\epsilon}{M} \;, \nonumber \\
\dot{\epsilon} &=& \left( -M \Omega^2 + \frac{2\gamma_E N_b}{\lambda^2} 
- \frac{4\gamma_E N_q}{\lambda^2} \right) \eta
\;.
\label{FPVI}
\end{eqnarray}
The corresponding eigenvalues become imaginary under the following inequality
\begin{equation}
\frac{2\gamma_E}{\lambda^2} \left(N_b-2N_{q}\right) <  M\Omega^2 \;,
\end{equation}
which leads to stable, oscillatory dynamics around the origin. In other words, satisfying the inequality 
specifies the parameters for a single-well potential. Note that $N_q$ has an inverse square dependence on 
$\lambda$ so the boundary equality involves a quadratic in $\lambda^2$.  Alternatively, one can numerically 
find the fixed point value for $Q^*$ by turning Eq.(19) into an iterative map. We will discuss this later 
after first completing the analysis for the velocity-dependent form of the interaction.

\begin{figure*}[t]
\begin{center}
\includegraphics[width=0.48\textwidth]{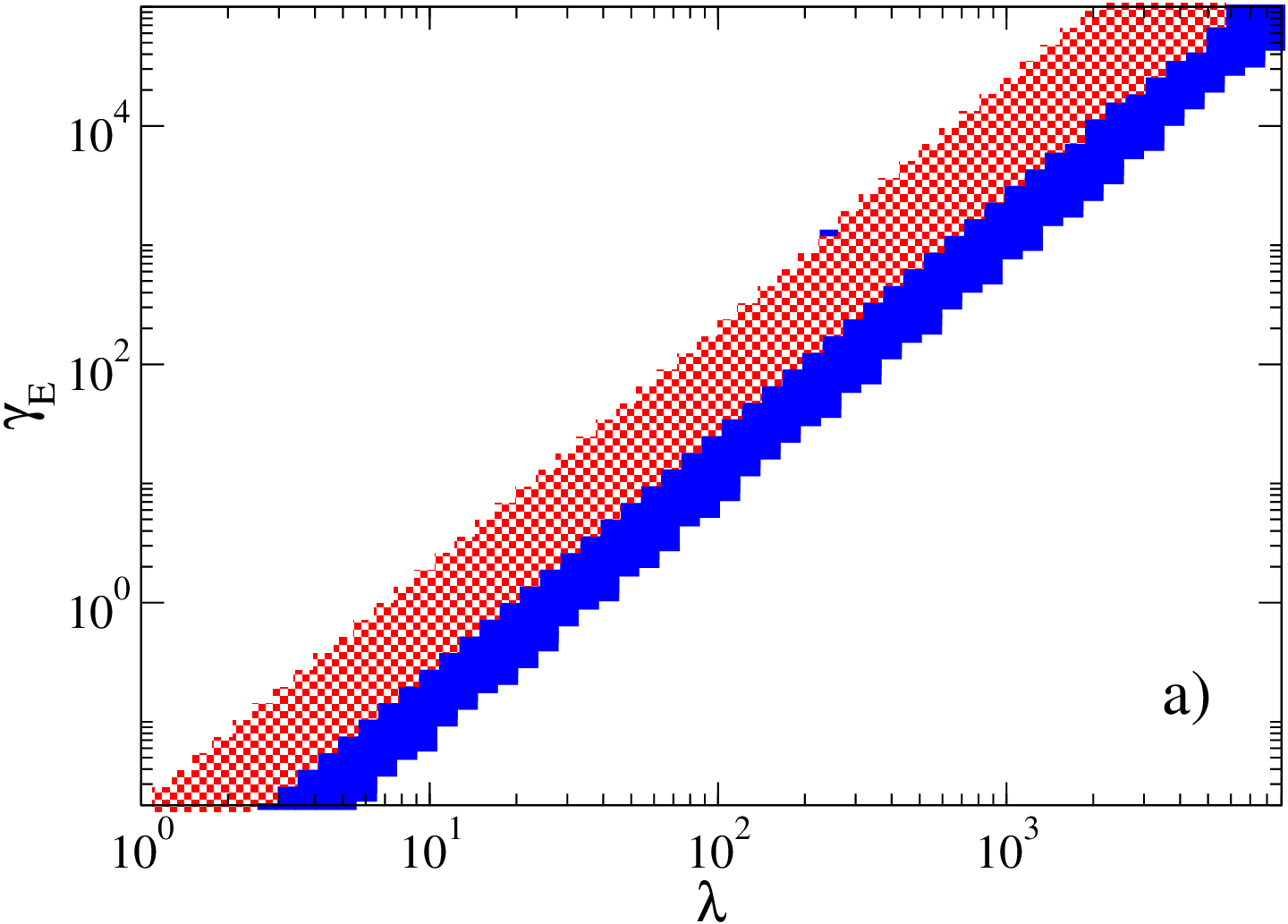}
\hspace{0.2in}
\includegraphics[width=0.48\textwidth]{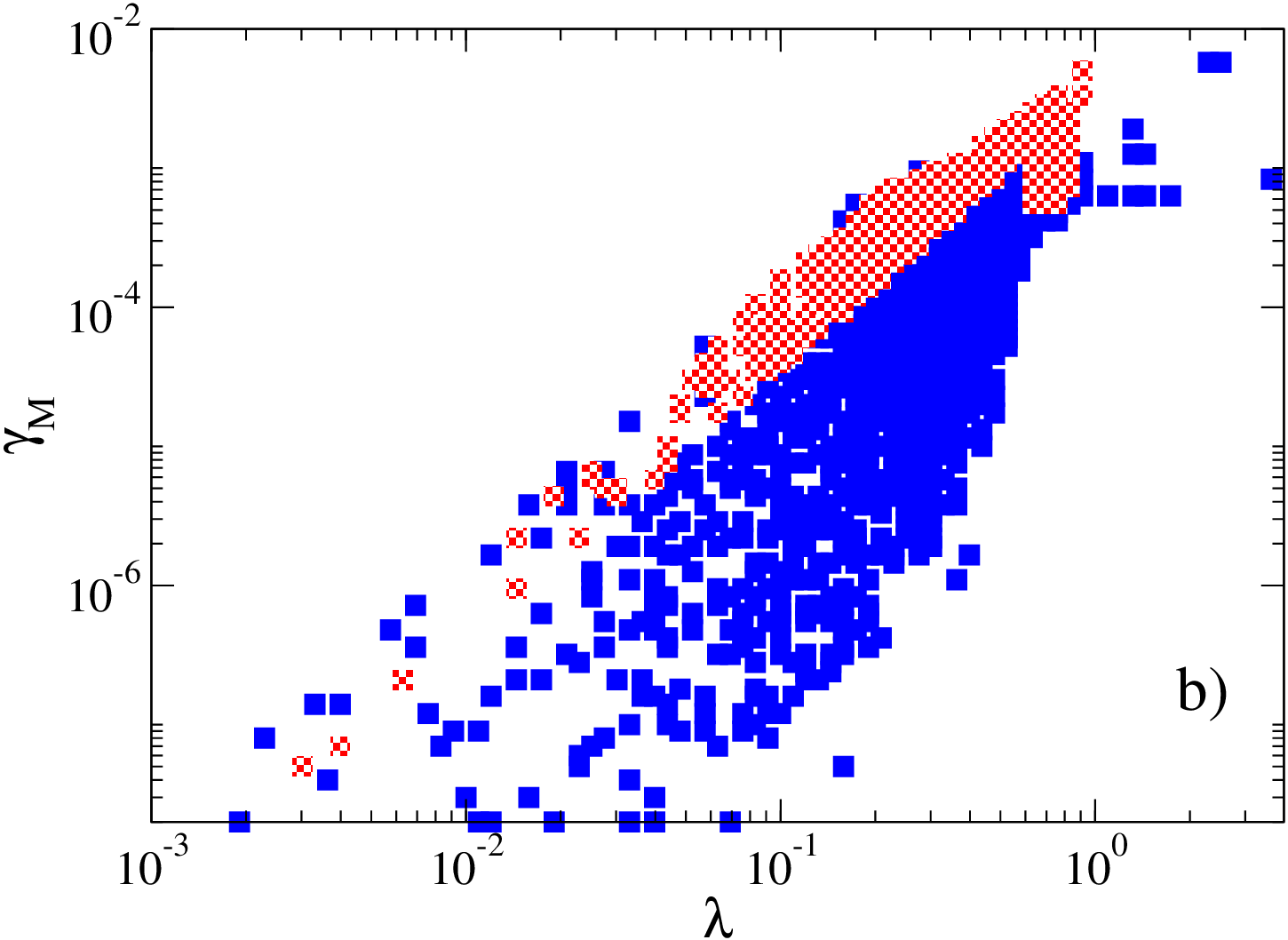}
\end{center}
\caption{(Color online) Stability diagrams in parameter space for the 
dynamical fixed point at $(P^*,Q^*)=(0,0)$ for the two forms of the interaction Hamiltonians, 
velocity-independent (a) ($\gamma_E,\lambda$ parameter space) and velocity-dependent (b) 
($\gamma_M,\lambda$ parameter space). The white background denotes values where the origin 
is stable (effective single-well potential). The blue region corresponds to fixed points off the origin 
(due to the effective double-well dynamics), while the red (chessboard-shaded) region is a boundary where 
our iterative scheme does not converge to within $\epsilon=10^{-10}$, which may be viewed as a 
critical boundary. Other relevant parameters in both cases shown are $N_b=10^3$, $m=M=1$, $\omega_n=1$, 
$\Omega=144/89$, and $T_b=0.5$. The convergence condition is set by $\epsilon$ and the maximum 
number of iterations is $10^4$. Masses, angular frequencies, temperatures, and parameters $\gamma_E$, 
$\gamma_M$, and $\lambda$ are expressed in arbitrary units.}
\label{Fig2}
\end{figure*}

\subsection{Velocity-dependent interaction}
 
By repeating analogous considerations for the velocity-dependent case, the equilibrium location 
$(Q^*, P^*)$ can be shown to satisfy the conditions
\begin{eqnarray}
\frac{P^*}{M} &=& \frac{2\gamma_M}{M} \sum_{n=1}^{N_b} \left(\frac{p_n}{m} - 
\frac{P^*}{M}\right) \exp{[-(q_n-Q^*)^2/\lambda^2]} \;, \\
M\Omega^2 Q^* &=& - \frac{2\gamma_M}{\lambda^2} \sum_{n=1}^{N_b} \left(\frac{p_n}{m} - 
\frac{P^*}{M} \right)^2 (q_n-Q^*) \exp{[-(q_n-Q^*)^2/\lambda^2]} \;.
\end{eqnarray}

In the case of small $\lambda$, $Q^*=0$ and Eq. (22) reduces to
\begin{equation}
\frac{P^*}{M} = -\frac{2\gamma_M}{M^2}  P^*\;,
\end{equation}
modulo a multiplicative constant on the righthand side, which also means $P^*=0$. Thus here again 
the origin in phase space, $(0,0)$, is the only equilibrium point in the case of small $\lambda$. 

The opposite limit of large $\lambda$ can be treated again with the expansion of the exponential 
which leads to
\begin{eqnarray*}
\frac{P^*}{M} &=& \frac{2\gamma_M}{M} \sum_n \left(\frac{p_n}{m} - \frac{P^*}{M}\right) 
\left[ 1 - \frac{(q_n-Q^*)^2}{\lambda^2} \right] \ \\
&=& -\frac{2\gamma_M N_b}{M^2} P^* - \frac{2\gamma_M}{M} \sum_n \left(\frac{p_n}{m} - 
\frac{P^*}{M}\right) \frac{q_n^2-2q_n Q^*+(Q^*)^2}{\lambda^2} \ \\
&=& -\frac{2\gamma_M N_b}{M^2} P^* + \frac{2\gamma_M C_2}{M^2} P^* + 
\frac{ 2\gamma_M C_3}{M^2 \lambda^2} P^* (Q^*)^2 \;,
\end{eqnarray*}
where we have used the fact that $\sum_n p_n^\alpha q_n^\beta$ is assumed to be nonzero only when $\alpha$ 
and $\beta$ are both even, i.e. a symmetric probability density function in both $p_n$ and $q_n$, and 
the constants $c_i$ implicitly defined above are largely unimportant for our qualitative considerations. 
What is clear from the last step is that $P^*=0$ is still the equilibrium momentum value, and this 
implies conditions on $Q^*$ 
\begin{eqnarray*}
M\Omega^2 Q^* &=& -\frac{2\gamma_M}{\lambda^2} \sum_n \frac{p_n^2}{m^2} 
\left[1 - \frac{(q_n-Q^*)^2}{\lambda^2} \right](q_n-Q^*) \ \\
&=& - \frac{2\gamma_M}{\lambda^2} \sum_n \frac{p_n^2}{m^2} (q_n-Q^*) 
+ \frac{2\gamma_M}{\lambda^4} \sum_n \frac{p_n^2}{m^2} (q_n-Q^*)^3 \ \\
&=& + \frac{2\gamma_M d_1}{\lambda^2 m^2} Q^*- \frac{6\gamma_M d_2}{\lambda^4 m^2} Q^* - 
\frac{2 \gamma_M d_1}{\lambda^4 m^2} (Q^*)^3 \;,
\end{eqnarray*}
where $d_1 = \sum_n p_n^2$ and $d_2 = \sum_n p_n^2 q_n^2$. The functional form of this relation is 
$A(Q^*)^3 = B Q^*$ which means that the possible solutions are $Q^*=0$ and $Q^* = \pm \sqrt{B/A}$. 

Since even in this case $(0,0)$ is always a fixed point, a stability analysis around this phase space 
location analogous to the previous case should be sufficient to identify the threshold for bistable 
behavior, obtaining the linearized relations
\begin{eqnarray}
\dot{\eta} &=& \left( \frac{1}{M}+ \frac{2\gamma_M c_0}{M^2} \right) \epsilon \equiv  b \epsilon 
\;, \nonumber \\
\dot{\epsilon} &=& \left( - M\Omega^2 + \frac{2\gamma_M c_p}{m^2\lambda^2}  - 
\frac{4\gamma_M c_{qp}}{m^2 \lambda^2} \right) \eta \equiv a \eta
\;,
\end{eqnarray}
where
\begin{equation}
c_0 = \sum_{n=1}^{N_b} \exp{(-q_n^2/\lambda^2)},\; \;
c_p =\sum_{n=1}^{N_b} p_n^2 \exp{(-q_n^2/\lambda^2)}, \; \; 
c_{qp} = \sum_{n=1}^{N_b} p_n^2 (q_n/\lambda)^2 \exp{(-q_n^2/\lambda^2)} \;.
\end{equation}
The eigenvalues $\Lambda$ of the Jacobian matrix satisfy the condition $\Lambda^2 = ab$ which given that
$b>0$ (for $\gamma_M > 0$) results in $a<0$ in order for the eigenvalues to be imaginary indicating 
stability. This leads to
\begin{equation}
\frac{2\gamma_M}{m^2\lambda^2} \left(c_p-2c_{qp}\right) <  M\Omega^2 \;,
\end{equation} 
The other case, for fixed $\gamma_M$, is more involved. Note that the both terms on the left of the 
inequality vary with time, with the evolution of the bath variables, though each can be replaced by 
the averages over the bath multiplied by $N_b$. 

As indicated earlier, away from these limiting cases, we can use an iterative scheme to numerically 
determine the location of the fixed point. This is made easier by the fact that $P^*=0$ for both forms 
of the interaction and only $Q^*$ has to be determined. A wide range of parameter values can be considered 
and for each pair of $(\lambda,\gamma_E)$ or $(\lambda,\gamma_M)$ values, the location of the fixed point 
numerically computed. As depicted in Fig.\ref{Fig2}(a) and Fig.\ref{Fig2}(b) ,for velocity-independent 
and velocity-dependent cases respectively, the parameter values where the origin is a globally stable 
fixed point can be distinguished from those where the fixed point moves away. 
These phase diagrams can prove useful in determining regimes where the relaxation between the test 
particles and bath occurs more readily. It is worth noting the existence of a critical boundary 
layer where convergence slows down considerably, denoted by the red points in both panels in Fig. 2. 
The precise implications of this region is still to be determined though we suspect that this may 
simply be the consequence of the merger of the two wells (of the double-well potential) in becoming 
a single well.

\section{Thermalization dynamics: Numerical Results}

Having gained some insight into the behavior of a single target particle interacting with the oscillators
in the bath, we now move to the issue of thermalization. In our numerical exploration, we considered the 
distinct thermal relaxation scenarios of a single target oscillator interacting with the $N_b$ 
bath oscillators as well the more realistic situation where $N_{p}$ target oscillators interact with the bath
but the results presented here focus on the latter case. 
The equations of motion were integrated using a variable step algorithm which
preserved the Hamiltonian to machine precision. When multiple target oscillators were considered, the 
initial conditions were drawn from a thermal distribution with temperature $T_{p}$. 
Again, with an eye on the sympathetic cooling application, the temperature of the target particles 
was chosen to be higher than that of the bath, that is $T_p > T_b$. In other words, the species to be cooled
are the target particles with the coolant species making up the bath.
All the oscillators in each category were assigned the same angular frequency, $\omega$ 
for bath oscillators and $\Omega$ for the target ones. In light of earlier analysis within the 
Caldeira-Leggett framework~\cite{Taylor} and the conditions for sympathetic cooling of mixtures, we 
consider $\omega \neq \Omega$. Our choice of $\Omega=144/89$ is intended to preclude any low-order
resonances between bath and target particles though, as discussed a little later, this should not 
be a problem. The two other parameters of relevance are the range of the interaction $\lambda$ and 
their interaction strengths $\gamma_E$ (velocity-independent) and $\gamma_M$ (velocity-dependent). 
Although we numerically explored a range of model parameters, we have decided to limit the results shown 
to some illustrative cases which highlight new phenomena as well as being pertinent to application to 
atomic mixtures. One point to note is that we sample the dynamics every $\Delta t=0.01$ and our units 
of time are simply the number of time steps. This also means that a single period of a bath oscillator 
is about $628$ time steps, which should be treated simply as a reference rather than a relevant timescale. 
We will return to this issue later when discussing our results.

\begin{figure}[t]
\begin{center}
\includegraphics[width=1.0\textwidth]{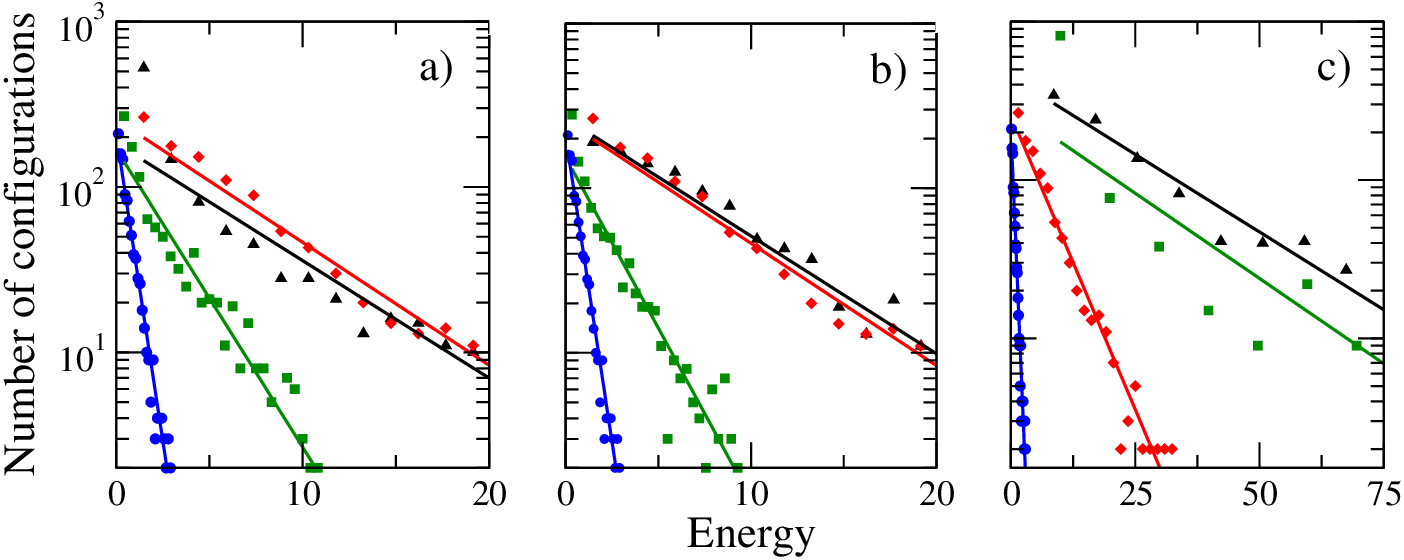}
\end{center}
\caption{(Color online) Energy spectrum of initial target system (red, diamond), 
bath (blue, circle) and final target (black, triangle), bath (green, square) for 
(a),(b) velocity independent and (c) velocity dependent interactions. 
In each instance, $N_b=N_{p}=10^3$ and initial temperatures set in the numerical 
code equal to $T_b=0.5$ for the bath, and $T_{p}=5$ for the target system 
(corresponding to inverse temperatures $\beta_b=T_b^{-1}=2$ and $\beta_{p}=T_p^{-1}=0.2$, 
respectively), $\omega=1.0, \Omega=144/89$. 
The parameters of the interacting Hamiltonian and the running time of the simulations are
(a) $\gamma_E=0.1, \lambda=0.01, t_{\mathrm{run}}=5 \times 10^4$,  
(b) $\gamma_E=0.1, \lambda=0.1,  t_{\mathrm{run}}=2 \times 10^4$, and 
(c) $\gamma_M=0.1, \lambda=0.1,  t_{\mathrm{run}}=10^4$.
The unweighted best fit to the Boltzmann distributions, shown as continuous lines around the 
points of each curve, provides  
$\beta_b^{(i)}=1.71 \pm 0.06$ and $\beta_t^{(i)}=0.171 \pm 0.009$ 
for the actual initial bath and target system inverse temperatures, respectively, and 
(a) $\beta_b^{(f)}=0.41  \pm 0.03 $,   $\beta_{p}^{(f)}=0.16 \pm  0.01 $, 
(b) $\beta_b^{(f)}=0.48  \pm 0.03 $,   $\beta_{p}^{(f)}=0.165 \pm 0.007$, 
(c) $\beta_b^{(f)}=0.049 \pm 0.009$,   $\beta_{p}^{(f)}=0.045 \pm 0.003$, for the final temperatures. 
Masses, angular frequencies, temperatures, parameters $\gamma_E$, $\gamma_M$, $\lambda$,  
energies and inverse temperatures are expressed in arbitrary units.}
\label{Fig3}
\end{figure}

Since we are primarily interested in the thermalization effect, all the cases discussed from 
now on involve multiple, $N_p$, target particles interacting with $N_b$ bath oscillators. 
Here again, we considered both velocity-independent and velocity-dependent forms of 
the interaction though there were very few qualitative differences between the behavior 
seen in the two instances. The initial conditions for both subsystems are drawn from 
thermal distributions and we considered a range of interaction times. 
The interaction parameters $\lambda,\gamma_E$ and $\lambda,\gamma_M$ have been chosen 
to correspond to situations where the fixed point at the origin in phase space is stable. The 
nonlinear regime discussed in Sections II and III would suggest a more favorable thermalization 
dynamics in the nonlinear regime when the fixed points are away from the origin in phase space.
However, the lack of overlap in the configuration space and the 
local form of the interaction makes this regime uninteresting for the task of a fast thermalization.
The energy or spectral distribution of each subsystem was then considered to extract the final 
temperature (actually inverse temperature) of both target particles and bath oscillators. 
It should be noted here that our use of the term 'temperature' is, in general, inexact given 
the dynamical and, hence, inherently non-equilibrium nature of the problem. 
Nevertheless, the slope obtained from fitting the logarithm of the energy distribution can 
be thought of as the effective temperature at a given time, with the error in its determination 
assessing also the effectiveness of the description in terms of this parameter alone. 
At any given time, we consider the total energy of the test particles, the total energy of the 
bath particles, and their interaction energy, defined as 
\begin{eqnarray}
E_{\mathrm{p}}(t) &=& \sum_{m=1}^{N_p} E_{m,\mathrm{p}}(t) = \sum_{m=1}^{N_{p}} 
\left(\frac{P_m^2(t)}{2M}+\frac{1}{2}M\Omega^2 Q_m^2(t)\right)\;, \\
E_{\mathrm{b}}(t) &=& \sum_{n=1}^{N_b} E_{n,\mathrm{b}}(t)= \sum_{n=1}^{N_b}    
\left(\frac{p_n^2(t)}{2m}+\frac{1}{2}m\omega^2 q_n^2(t)\right)\;, \\
E_{\mathrm{int}}(t) &=&\sum_{m=1}^{N_{p}} \sum_{n=1}^{N_b} H_{\mathrm{int}}(Q_m(t), P_m(t), q_n(t), p_n(t))\;. 
\end{eqnarray}
The distribution of energies for test and bath particles after a fixed interaction time $t_{run}$, 
that is the histograms of $E_{m,\mathrm{p}}(t_{run})$ and $E_{n,\mathrm{b}}(t_{run})$  respectively, can then be 
used to extract the effective inverse temperatures for the two subsystems. 
As we will discuss later, the energy content in the interaction also appears to play an important 
role in determining the final temperatures in both our models.

\begin{figure}[t]
\begin{center}
\includegraphics[width=0.48\textwidth]{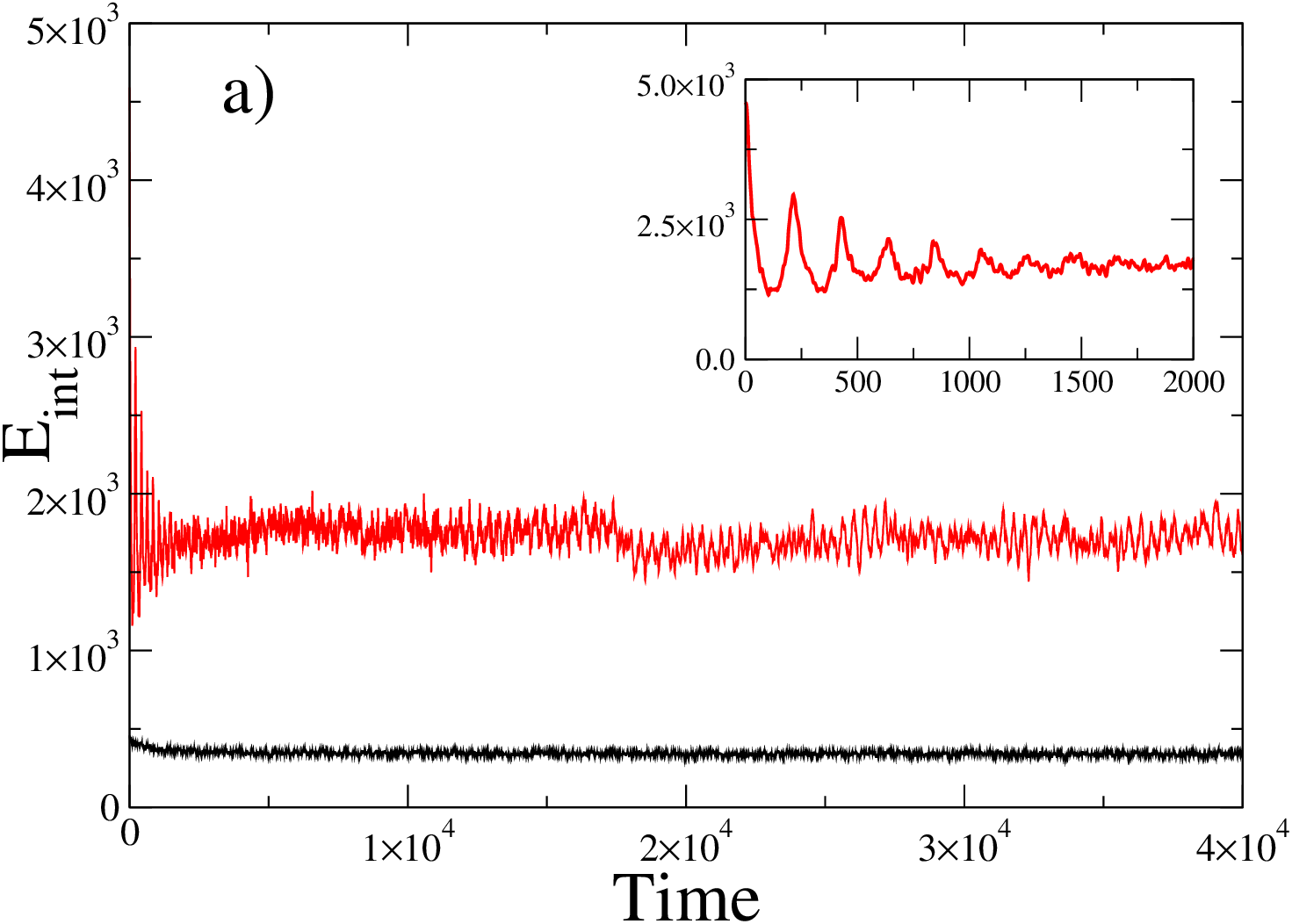}
\hspace{0.2in}
\includegraphics[width=0.48\textwidth]{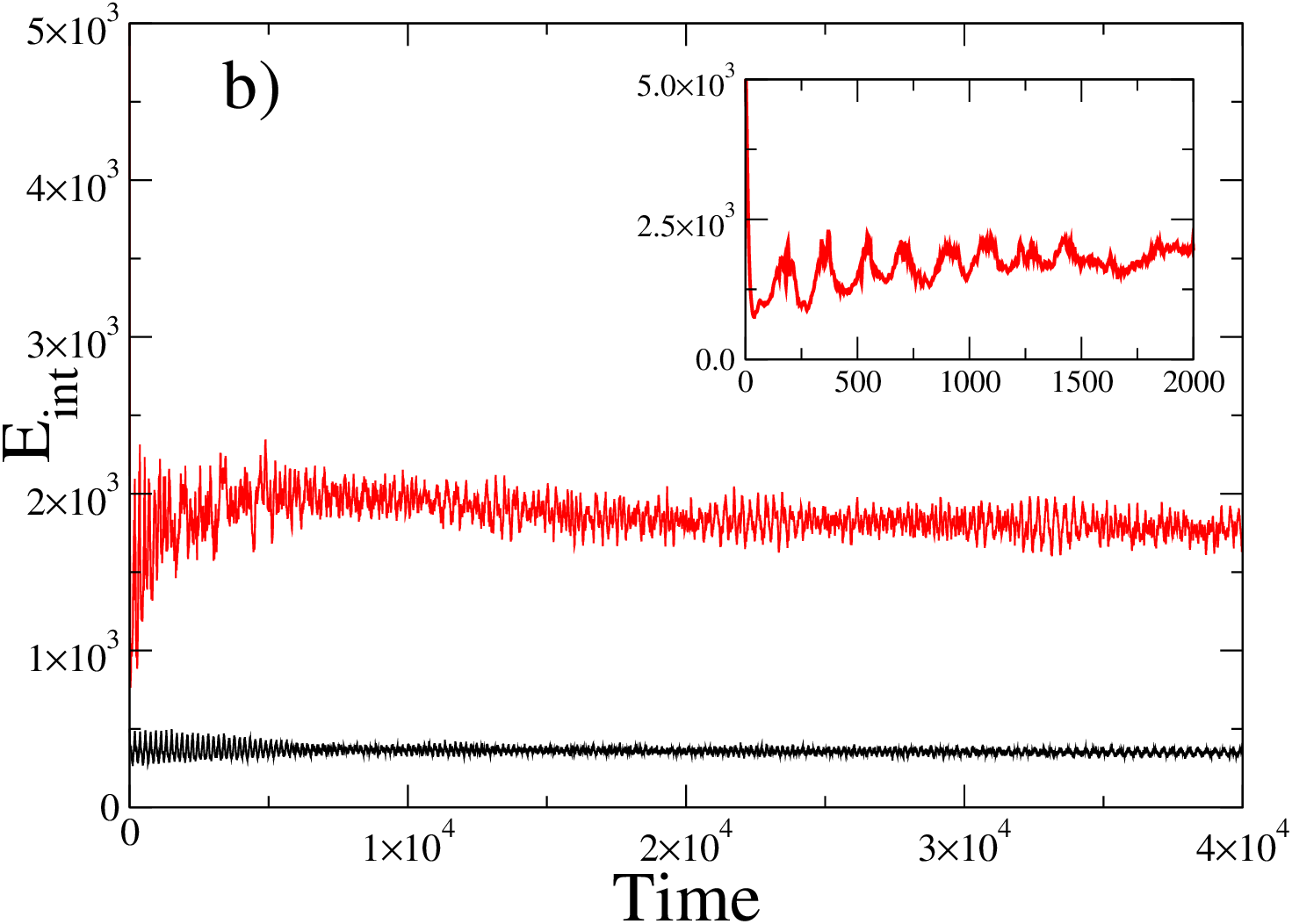}
\end{center}
\caption{(Color online) Interaction energy as a function of time for velocity independent (a) and 
velocity dependent (b) interactions. The plots are for two choices of $\lambda=0.01$ (black, bottom curve) 
and $\lambda=0.1$ (red, top curve), with $\gamma_E=0.1$ (a) and $\gamma_M=0.02$ (b), where the latter 
parameter was intentionally chosen to have comparable final interaction energy in the two situations. 
Oscillations which appear on short timescales are shown in the insets. All other parameters are 
chosen as in Fig. 3, and both time (in steps) and energies are expressed in arbitrary units.}
\label{Fig4}
\end{figure}

Figure~\ref{Fig3} shows the results of this construction for some illustrative cases where the  
lines drawn are least squares fits to the points which display the energy distributions at 
the times specified in the captions. In each case, both initial and final distributions of the 
two subsystems, particle and bath, are shown. The slope of the fit provides the corresponding 
$\beta=1/(k_B T)$ at time $t_{run}$.  Given the poor statistics at the higher ends of 
the energy spectra, we have also implemented weighted least squares fits 
which appear to be more accurate for any quantitative analysis though the qualitative 
inferences are rather robust with respect to the data fitting. The results in Fig.~\ref{Fig3} readily 
suggest the role the effective number of bath oscillators interacting with each target particle, 
$N_\lambda$, plays in promoting thermalization. Contrasting  panels (a) and (b) of Fig.~\ref{Fig3}, the 
interaction range $\lambda$ has been increased but the time can then been decreased to get the 
effective temperatures to be the same. Plot (c) in Fig.~\ref{Fig3} shows the effect of the velocity 
dependent interaction at shorter times, and implies temperature oscillations in the early stage of 
thermalization. The oscillatory phenomenon is more clearly seen by considering the time dependence of 
the interaction energy, defined as $E_{\mathrm{int}}=\gamma_E \sum_{n=1}^{N_b} \sum_{m=1}^{N_{p}} 
\exp{[-(q_n-Q_m)^2/\lambda^2]}$ in the case of velocity-independent interactions. 
As shown in the panels of Fig.~\ref{Fig4}, for both interaction types, there is a sharp decrease 
in this measure to essentially a quasi-stationary value, with the inset emphasizing the energy 
oscillations at early times. The velocity-dependent case shows more quasi-periodicity than in 
the velocity-independent situation. 

\begin{figure*}[t]
\begin{center}
\includegraphics[width=1.0\textwidth]{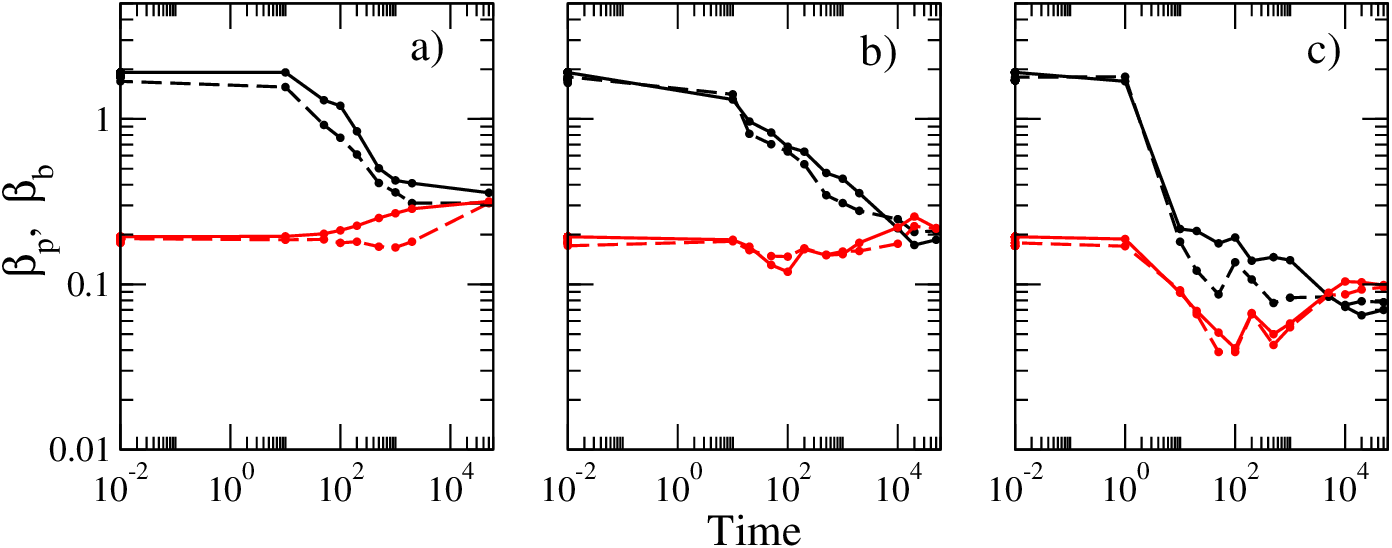}
\end{center}
\caption{(Color online) Relaxation dynamics for target and bath subsystems for velocity independent 
interactions. The effective inverse temperature is shown versus time for the test particles with unweighted 
(red dashed line) and weighted (red continuous line) fitting to a Boltzmann distribution, bottom curves on 
the left side of each plot, and for the bath particles with unweighted (black dashed line) and weighted 
(back continuous line) fitting, top curves on the left side of each plot. 
In each instance, $N_b=N_{p}=10^3$, $T_b=0.5$, $T_{p}=5.0$, $\omega=1.0$, $\Omega=144/89$, with 
interaction parameters (a) $\gamma_E=0.1, \lambda=0.01$; (b) $\gamma_E=0.1, \lambda=0.1$, and 
(c) $\gamma_E=0.5, \lambda=0.1$. While the weighted fits are expected to be more accurate, significant 
deviations from the results of unweighted fits can be used to infer the presence of strong deviations
from a Boltzmann distribution. The spread in the left most data points is a consequence of small 
statistics associated with the sampling of the thermal initial conditions through different seeds 
for the random number generator. The typical relative errors associated with the inverse temperature 
determination through the fits are about $10 \%$, and masses, angular frequencies, parameters $\gamma_E$, 
$\lambda$, energies, and inverse temperatures are expressed in arbitrary units.}
\label{Fig5}
\end{figure*}

\begin{figure*}[t]
\begin{center}
\includegraphics[width=1.0\textwidth]{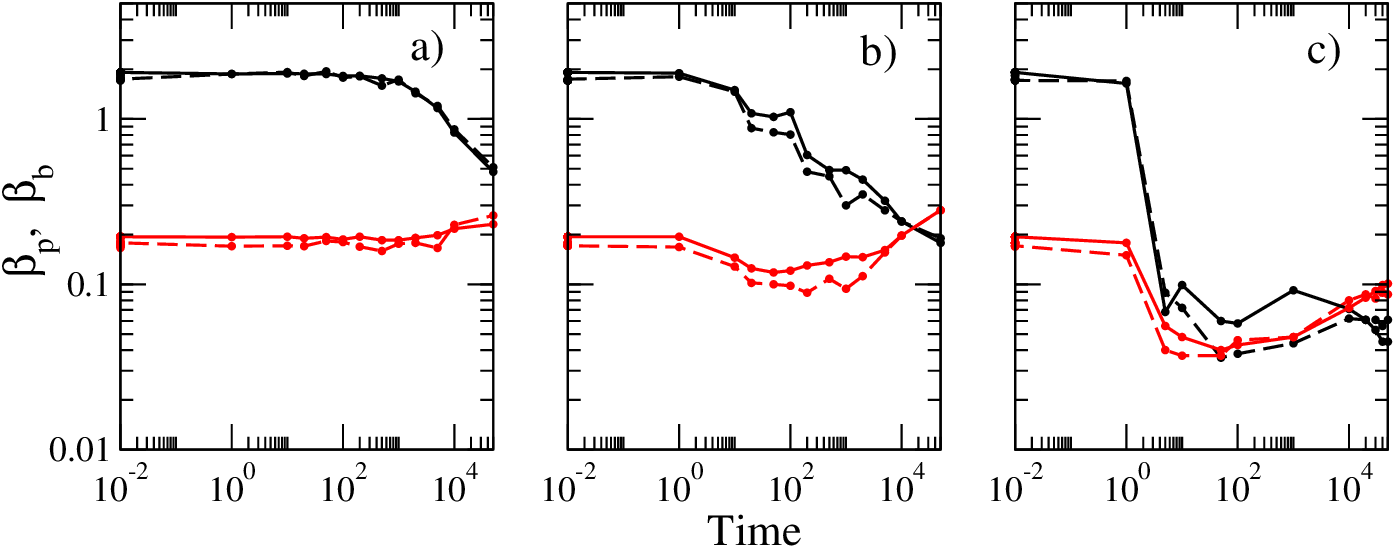}
\end{center}
\caption{(Color online) Relaxation dynamics for target and bath subsystems for velocity dependent 
interactions. The effective inverse temperature is shown versus time for the test particles with 
unweighted (red dashed line) and weighted (red continuous line) fitting to a Boltzmann distribution, 
bottom curves on the left side of each plot, and for the bath particles with unweighted (black dashed line) 
and weighted (black continuous line) fitting, top curves on the left side of each plot. 
Like in Fig. 5, we have $N_b=N_{p}=10^3$, $T_b=0.5$, $T_{p}=5.0$, $\omega=1.0$, $\Omega=144/89$, while 
the interaction parameters are (a) $\gamma_M=0.02, \lambda=0.01$; (b) $\gamma_M=0.02, \lambda=0.1$ and (c) 
$\gamma_M=0.1, \lambda=0.1$. The spread in the left most data points is a consequence of small statistics 
associated with the sampling of the thermal initial conditions through different seeds for the random 
number generator. The typical relative errors associated with the inverse temperature determination 
through the fits are about 10 $\%$, and masses, angular frequencies, parameters $\gamma_M$, 
$\lambda$, energies, and inverse temperatures are expressed in arbitrary units.}
\label{Fig6}
\end{figure*}

The observant reader will also notice that in Fig. 3(c) the final inverse temperatures of both bath 
and target subsystems are lower than the initial values, implying heating of both species.
To explore the dynamics of thermalization  more clearly, and in particular this double heating 
effect, we consider the time dependence of the effective temperatures of bath and target subsystems.
Based on what we deduced earlier, thermalization should occur faster in the case with larger $\lambda$, 
irrespective of the nature of the interaction, and relatively quicker for the velocity-dependent interaction. 
The latter expectation is trickier to show numerically as equivalent parameter regimes for the two cases 
are hard to determine. This is readily seen from the parameter phase diagrams shown earlier. 
However, as seen from contrasting Figs.~\ref{Fig5} and \ref{Fig6}, where three cases for each 
interaction type is shown, the qualititative features are very similar. In each figure, what is shown 
is (a) where thermalization does not occur even at the longest time considered; 
(b) where there are clear signs that thermalization is occurring and 
(c) an anomalous situation where the final temperatures for both bath and target are higher 
than their initial values. Figures~\ref{Fig5}(a) and (b) (and their equivalents in Fig.~\ref{Fig6}) 
clearly show that the temporal scale for relaxation decreases on increasing the spatial region of interaction 
(increasing $\lambda$).  This in turn implies faster equilibration due to the increase in the
average number of interacting particles at any given time.  

We now turn to the anomalous situation seen in both Figs.~\ref{Fig5}(c) and ~\ref{Fig6}(c). 
This occurs, for certain initial configurations, when a large amount of interaction energy is 
initially present in the system, typically in the large $\gamma_E$ or $\gamma_M$ limit where 
we will have $E_{\mathrm{int}}>>E_b, E_p$. This is a distinctive feature of our approach, as usually any 
initial interaction energy or  initial correlation between two subsystems is assumed to be 
nearly negligible with respect to the internal energy of the subsystems. 
This last situation is however characteristic of solid state systems, in which 
the contact between two bodies is limited to their surfaces, and it is therefore 
marginal with respect to the internal energy of the bodies. 
Here instead the two 'bodies' are penetrating, as atoms of the two different 
species see each other in the whole available trapping volume. 
Clearly this is outside the usual view of systems in thermal contact where the interface 
plays no role in the thermalization process. Here it does and, given the excess energy content, 
can heat both the systems as if there is a fictitious third, hotter subsystem in the problem, or 
if the interaction energy acts as a sort of 'latent heat' released during the time evolution. 
Of course this initial interaction energy can also be minimized, even at a relatively large 
value of $\gamma_E$ or $\gamma_M$, by tactically choosing the initial conditions in such a 
way that the oscillators of the two systems are very close to each other (in the case 
of attractive interactions) or very far apart (in the case of repulsive interactions). 
Likewise, the sudden change in the interaction between the two subsystems, for instance by exploiting  
Feshbach resonances of the interspecies scattering length, allows for implosions or explosions 
of the atomic clouds, as in the 'Bosenova' effect \cite{Saito,Donley,Altin}. All these situations 
are in principle covered by our simulation technique by properly choosing initial conditions for 
the bath particles and/or time-dependent interactions couplings.

Regarding the thermalization timescale and its dependence upon the parameters $\gamma_E$, 
$\gamma_M$, and $\lambda$, the plots in Figs.~\ref{Fig5} and \ref{Fig6} show that the 
thermalization time is inversely proportional to the interaction strength as intuitively expected. 
This can be confirmed by a semiquantitative analysis at least in the simpler case of velocity 
independent interactions. By visual inspection of the right hand side of Eq. (7), the two relevant 
timescales for the test particle are the intrinsic one of its free oscillation (the period of 
oscillation $T=2\pi/\Omega$) and a {\sl response} time to the force due to any single oscillator 
of the bath, $\tau_r \simeq 2 \pi [M\lambda^2/(2\gamma_E)]^{1/2}$. 
Thus, the thermalization timescale, proportional to $\tau_r$ varies with the coupling strength 
as $\propto 1/\sqrt{\gamma_E}$. Indeed, in Fig.~\ref{Fig5}(b) and (c), corresponding 
to the same $\lambda$ with coupling strengths $\gamma_E$ differing by a factor 5, the crossing between 
the two curves corresponding to the weighted fits for the inverse temperature occurs at times 
in the approximate ratio of 2 ($\simeq 10^4$ time steps for the case in (b), $\simeq 5 \times 10^3$ 
times steps in (c)). A similar analysis is unfortunately not possible for the dependence of the 
thermalization time upon the interaction range $\lambda$. A naive extension of the argument above 
indicates a response time increasing linearly with $\lambda$, which defies physical intuition 
as in this case more and more oscillators interact with the test particle. This, in fact, is precisely 
the issue as $\tau_r$ introduced above does not take into account the number of interacting oscillators 
from the bath, {\it i.e.} the sum term in Eq. (7) which also grows with $\lambda$. Numerically, 
Fig.~\ref{Fig5} (a) and (b) show that increasing $\lambda$ speeds up thermalization. Due to the complicated 
structure of the Hamilton equations for the velocity-dependent case, the above analysis cannot be 
repeated, however the results shown in Fig.~\ref{Fig6} confirms the same trends in this case.
Also, notice that the onset of a quasi-stationary regime for the interaction energy in Fig.~\ref{Fig4} 
appears to coincide with the thermalization timescales in Figs.~\ref{Fig5} and \ref{Fig6}, suggesting 
that the former may act as a shortcut to study the relaxation to thermal equilibrium or at least to 
steady states. Finally, we show in Fig.~7 the long time behavior in the velocity-independent 
case, both for the evolution of the inverse temperatures (left panel) as well as the interaction energy 
(right panel). We note that not only do the timescales for thermalization at very short times
coincide for these two observables, corroborating the remark made in the previous sentence, but also 
that their long-time behavior seem to exhibit fluctuations of similar relative amplitude. Further, Fig.~7(a)
confirms that the crossing of the inverse temperatures clearly visible at shorter times (in both Figs. 5(b) and (c))
is a temporary feature, leading to a stabilization of the temperatures for later times. 

\begin{figure}[t]
\begin{center}
\includegraphics[width=0.48\textwidth]{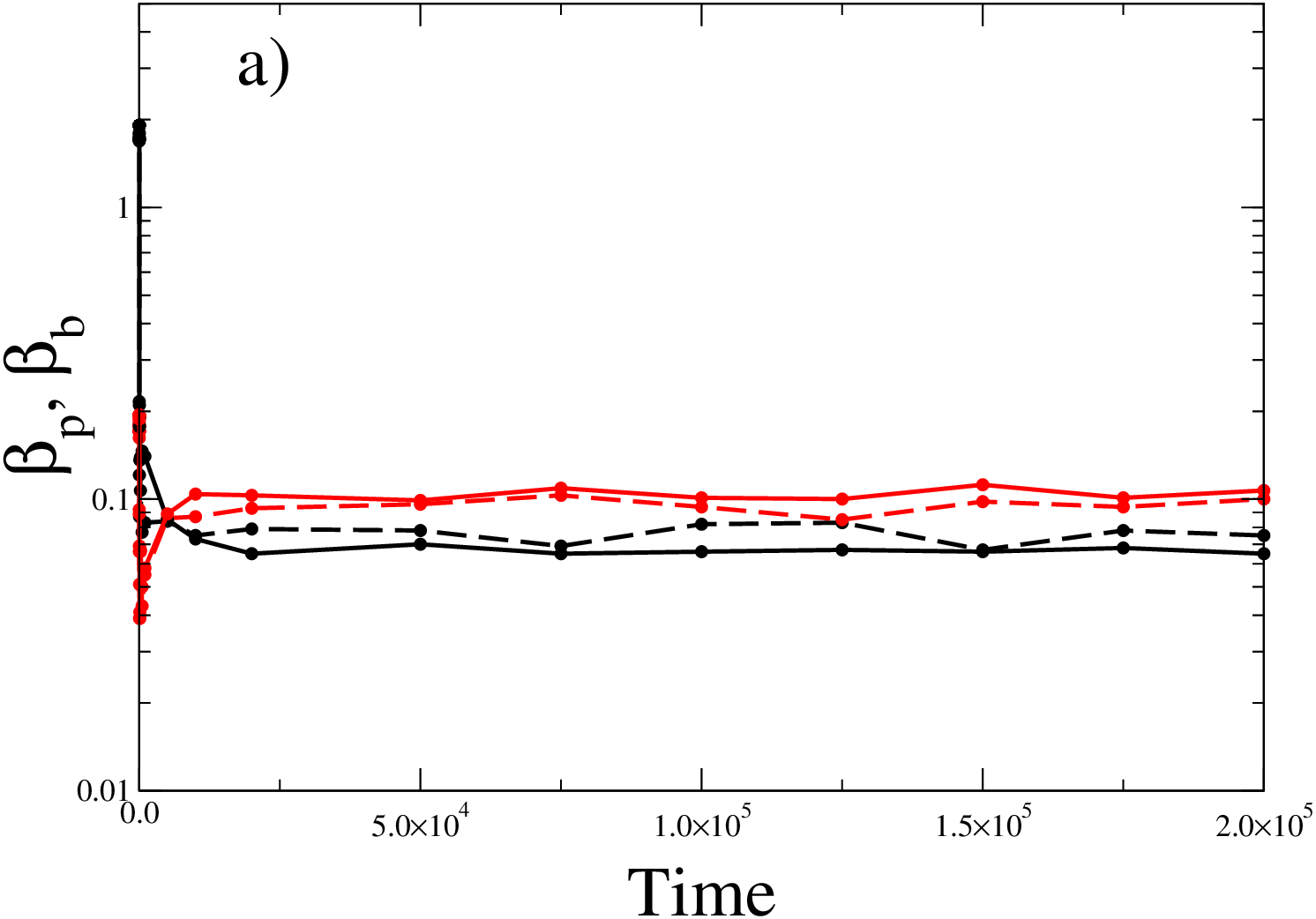}
\hspace{0.2in}
\includegraphics[width=0.48\textwidth]{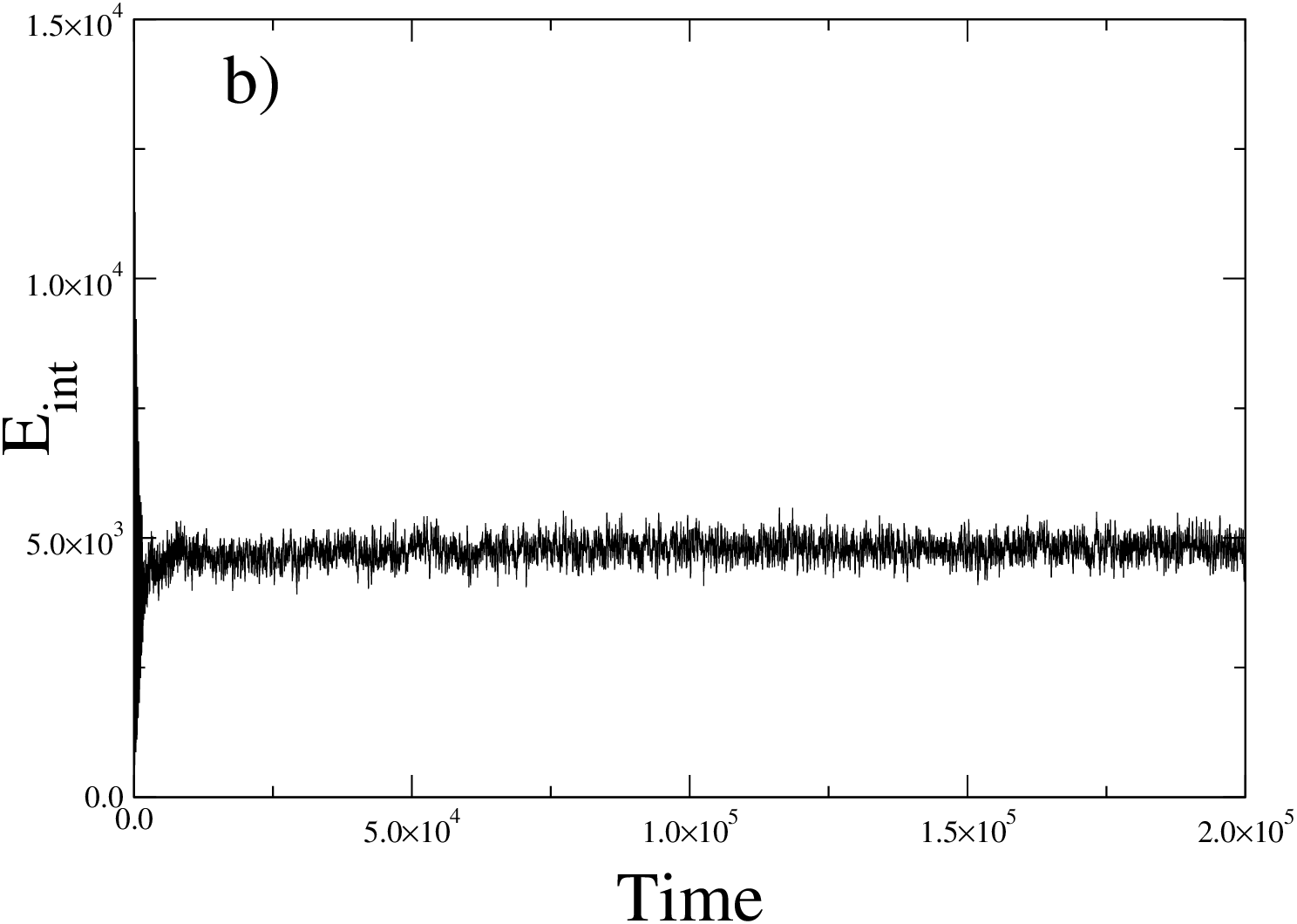}
\end{center}
\caption{(Color online) Long time simulation for the velocity-independent case. 
On the left plot (a) the effective inverse temperature is shown versus time for the test particles with 
unweighted (red dashed line) and weighted (red continuous line) fitting to a Boltzmann distribution, 
as in the two top curves on the right side of the plot, and for the bath particles with unweighted 
(black dashed line) and weighted (black continuous line) fitting, as in the two bottom curves on 
the right side of the plot. On the right plot (b) the corresponding interaction energy as a 
function of time is depicted. The parameters are chosen as in the case (c) of Fig. 5, and all 
quantities are expressed in arbitrary units.}
\label{Fig7}
\end{figure}

\section{Conclusions}

In conclusion, we have analyzed the thermalization dynamics of atoms trapped in a confining potential 
under the action of another gas with a generic temperature and different trapping frequency. 
The analysis has been carried out through molecular dynamics simulations adapting the 
Caldeira-Leggett model to the specific context of atoms oscillating at well-defined frequencies 
in a harmonic trap. The nonlinear form of the interaction between the two atomic species required 
an assessment of the dynamical stability which limits the range of the parameters of the model for
enabling efficient thermalization. Our numerical analysis does not rely on the weak-coupling 
assumption typically underlying the application of the Caldeira-Leggett model to open systems. 
A manifest implication of this more generalized framework is the fact that in the strong coupling 
regime the very definition of Boltzmann equilibrium distribution is at stake, and then only 
{\it effective} temperatures can be defined. This is in line with former analyses in the context 
of chemical physics, as in \cite{Montroll,Andersen1} in which the preservation of the Boltzmann 
distribution is ensured only in a weak interaction regime, and therefore fails for strongly 
exothermic chemical reactions (for a discussion of the strong coupling regime in open quantum 
systems, with similar outcomes in terms of lack of thermalization, see \cite{BeiLok}).

The simulations have been performed for a purely classical setting which, in terms of atomic 
clouds, seem justified whenever the sample temperature $T$ fulfils the condition $T >> \hbar \omega/k_B$,  
as usual in atomic trapping even at the lowest explored temperatures. Quantum effects will however 
influence the thermalization dynamics at finite temperature through the deviations from the classical 
Dulong-Petit specific heat law, which could be incorporated in the molecular dynamics code \cite{Dammak}. 
In particular, this affects the dynamics of thermalization for mixtures of Bose and Fermi gases
as their thermal response is quite different, with the specific heat scaling differently
with temperature as they become degenerate. Simulations in this setting should be able to confirm 
the existence of a heat capacity mismatch that strongly hinders the capability of a Bose gas to 
sympathetically cool a Fermi gas, as discussed in \cite{Truscott,Wouters,Presilla,BrownWei}. 
If the interaction strength between the two species is not large enough, the Fermi gas will not 
reach an equilibrium state with the Bose gas undergoing evaporative cooling, an issue of crucial 
relevance for precision thermometry of degenerate Fermi gases as emphasized in \cite{DeMarco}.
It is also worth remarking that, in the case of harmonic oscillators, classical dynamics and quantum 
dynamics of the centroid coincide, based on the Ehrenfest theorem, so there is decoupling between average 
motion and quantum fluctuations. This allows the analysis, even in a quantum framework, of the simpler case 
of a classical harmonic oscillator, with additional but separate considerations for the fluctuation
part of its dynamics in its quantum counterpart, provided that the nonlinear interaction term is 
perturbative with respect to the uncoupled dynamics. 

Various aspects of our model could be expanded in future work. Specifically, the 
role of the particle masses of the two systems in determining thermal 
equilibrium. This could allow the generalization of the notion of Rayleigh and Lorentz gases, corresponding 
to the two extreme limits of $m << M$ and $m >> M$ respectively, to arbitrary interaction potentials, 
not just restricted to the case of hard-sphere interactions discussed in Ref. ~\cite{Andersen2}. 
The two phenomenological parameters of the interaction need to be mapped to {\it ab initio} parameters 
of interatomic interactions in order to  promote quantitative analyses for concrete experimental settings. 
Further, by changing our choice of initial conditions, situations 
in which fast atoms with a narrow Gaussian energy distribution approach an equilibrium thermal 
cloud can also be recovered, bridging the experimental studies performed in~\cite{Zhang}.
The model could be also be applied to more general dynamical situations, for instance 
time-dependent interactions in an atomic mixture resulting from a sudden change of the interspecies 
scattering length or sympathetic cooling under targeted driving of the trapping frequencies, as in 
shortcuts to adiabaticity~\cite{Chen,Schaff,Torrontegui}. 
The latter situation could make more stringent the practical feasibility of the proposal 
introduced in Ref.~\cite{Choionba1}, or, more generally, the optimization of a targeted quantum 
state at finite temperature discussed in Ref.~\cite{Mukherjee}.

\end{document}